# A new solving procedure for the Kelvin–Kirchhoff equations in case of falling a rotating torus

**Sergey V. Ershkov\***, Plekhanov Russian University of Economics,
Scopus number 60030998, Department of Scientific Research, *Stremyanny* lane 36, Moscow 117997, Russia, e-mail: sergej-ershkov@yandex.ru,

**Dmytro Leshchenko**, Odessa State Academy of Civil Engineering and Architecture, Odessa 65029, Ukraine, e-mail: leshchenko_d@ukr.net

**Ayrat R. Giniyatullin,** Nizhny Novgorod State Technical University n.a. R.E. Alekseev, 24 Minina st., Nizhny Novgorod 603155, Russia, e-mail: araratishe@gmail.com

**Abstract**

We present in this communication a new solving procedure for Kelvin–Kirchhoff equations, considering the dynamics of falling the *rigid* rotating torus in an ideal incompressible fluid, assuming additionally the dynamical symmetry of rotation for the rotating body, $I_1 = I_2$.

Fundamental law of angular momentum conservation is used for the aforementioned solving procedure. The system of *Euler* equations for dynamics of torus rotation is explored in regard to the existence of an analytic way of presentation for the approximated solution (where we consider the case of laminar flow at slow regime of torus rotation). The second finding is associated with the fact that the Stokes boundary layer phenomenon on the boundaries of the torus is also been assumed at formulation of basic Kelvin-Kirchhoff equations (for which analytical expressions for the components of fluid's torque vector $\{T_2, T_3\}$ were obtained earlier). The results of calculations for the components of angular velocity $\{\Omega_i\}$ should then be used for full solving the momentum equation of Kelvin–Kirchhoff system. Trajectories of motion can be divided into, preferably, 3 classes: zigzagging, helical spiral motion, and *the chaotic regime* of oscillations.

# 1. **Introduction, equations of motion.**

Kelvin–Kirchhoff equations describe the dynamics of rigid body motion in an ideal fluid [Kirchhoff, 1877], [Ershkov, *et al.* 2020*a*], expressing the conservation of linear and angular momentum for the coupled fluid-to-body interaction problem (and *vice versa*).

The variability of the aforementioned motion observed in the trajectories during falling or ascending of particles (or of bodies) can be divided into, preferably, 3 classes of motions: zigzagging, helical spiral motion, and the chaotic regime of oscillations. We should mention the recent comprehensive works [Mathai, *et al.* 2018], [Pan, *et al.* 2019], [Ern, *et al.* 2012], [Novikov & Shmel'tser, 1981] and [Kozelkov, *et al.* 2018], illuminating this problem significantly (e.g., in [Kozelkov, *et al.* 2018] was considered the special case of particles rising in a *viscous* fluid).

In our first work [Ershkov, *et al.* 2020*a*] (from the cycle of works [Ershkov, *et al.* 2020*a*], [Ershkov, *et al.* 2020*b*]) absolutely new approach in procedure for solving Kelvin–Kirchhoff equations, for obtaining the elegant analytical and semi-analytical solutions to the system of ODEs (at describing the dynamics of rigid body's motion in an ideal fluid) was suggested in case of the ascending or falling the rigid sphere e.g. deep inside the Ocean.

In the current work we present the solution for the same problem (Kirchhoff equations) but having considered the dynamics of the rotating torus at falling in an ideal fluid.

Despite that the system of initial equations (1)-(2) below (and, hence, the general method of solving (3)-(6)) are the same as suggested in [Ershkov, *et al.* 2020*a*] earlier, the applied torques quite differ in simple case of sphere, presented in the work [Ershkov, *et al.* 2020*a*], and in case of falling the rotating torus.

Indeed, one can see similar formulations of the aforementioned problems, but the way of solving and results differ significantly if you compare them one to each other: the dynamics of rotations of rigid sphere (and, hence, its motion) is determined by only one but arbitrary torque component $T_3$; whereas, in simple case of a spinning torus (rotation of a torus about its axis of symmetry) one can see that analytical solving of Euler equations of angular momentum yields quite differing results (13)-(15) for the components of angular velocity of torus rotation, which should determine the quasi-periodic or even *chaotic* regime of oscillations during its motion in a fluid. But this is only a simple case of a spinning torus; meanwhile, motions of a slender torus have been classified into to 5 types (2 types of translational motion, 2 types of rotations as well as the case of expanding torus).

We consider here the case of rigid torus only, so the case of expanding torus should be excluded. All the types of motions (with appropriate components of torque) have been considered in the current research accordingly, as well as the regimes of torus rotation in a fluid (depending on the expressions for the components of torques on the surface of torus during its rotation). Moreover, we should especially note that the variability observed in the regimes of rotation during falling of body in a form of slim torus can be associated with two basic regimes of flows, laminar and turbulent, along with switching between them. It is shown that even at laminar regime of flows the chaotic regime during torus motion in a fluid may have arisen which stem from the quasi-periodic oscillations at non-linear torus rotation (17)-(19). This is the obvious basis for *chaotic* scenario at the falling of the rotating torus inside the fluid.

Let us outline here in the Section 1 the initial equations (1)-(2), suggested earlier in [Ershkov, *et al.* 2020a] (in addition, the general method of solving (3)-(6) let us outline in Section 2).

Most of researchers outline a key importance of the two governing parameters for the dynamics of rigid body's motion in an ideal fluid: 1) the rigid body's density $\rho_b$ relative to the density of fluid $\rho_f$ ($\Gamma \equiv \rho_b/\rho_f$, $\Gamma < 1$), and 2) its Galileo number Ga $\equiv \frac{1}{\nu}\sqrt{g D^3 (1-\Gamma)}$ ($g$ is the acceleration due to gravity, $D$ is the body's mean external rotational diameter, and $\nu$ is the kinematic viscosity of the fluid in the viscous Stokes boundary layer on the boundaries of the body).

According to approach used earlier [Ershkov, *et al.* 2020a], [Mathai, *et al.* 2018],

in the co-rotating frame of a Cartesian coordinate system $\vec{r} = \{x, y, z\}$, the Kelvin–Kirchhoff equations are the momentum equations (*at given initial or boundary conditions*):

$$\left(\Gamma + \frac{1}{2} + B_U \delta\right)\frac{d\vec{U}}{dt} + \Gamma \vec{\Omega} \times \vec{U} = \frac{\vec{F}}{m_f} + (\Gamma - 1)g, \Rightarrow$$

$$\frac{d\vec{U}}{dt} + A\vec{\Omega} \times \vec{U} = \vec{f}, \qquad (1)$$

$$A = \frac{\Gamma}{\Gamma + \frac{1}{2} + B_U \delta}, \quad \vec{f} \equiv \begin{Bmatrix} f_x \\ f_y \\ f_z \end{Bmatrix} \equiv \frac{\frac{\vec{F}}{m_f} + (\Gamma - 1)g}{\Gamma + \frac{1}{2} + B_U \delta}$$

which, nevertheless, should be additionally combined with the fundamental law of angular momentum conservation [Mathai, *et al.* 2018]

$$\frac{d\vec{K}}{dt} + [\vec{\Omega} \times \vec{K}] = \vec{T}, \qquad (2)$$

where in (1)-(2): $\Gamma$ is the body's mass-density ratio ($\Gamma \equiv \rho_b/\rho_f$, $\Gamma < 1$), $\vec{U}$ is the body velocity vector, $\vec{U} = \{v_1, v_2, v_3\}$; $g$ is the acceleration due to gravity; $\vec{K} = \{I_i \cdot \Omega_i\}$, whereas $\vec{\Omega} = \{\Omega_i\}$, here $\Omega_i$ are the components of angular velocity *pseudo*-vector $\vec{\Omega}$ along the principal axes, which determines the actual rate and direction of body's rotation, $i = 1, 2, 3$, and $I_i$ are the principal moments of inertia, in case of symmetric torus $I_1 = I_2$; $\vec{F}$ and $\vec{T}$ are the fluid's force and torque vectors, respectively, $m_f$ is the mass of the fluid displaced by the body. Note that $\delta = \sqrt{\frac{\nu \tau}{\pi D^2}}$ is the dimensionless Stokes boundary layer that develops in time $\tau$, and $D$ is the body's mean external rotational diameter ($\varepsilon = d/D$, where $d$ is the diameter of the torus's cross-section). The coefficients $B_U = 18$ is known analytically from the unsteady viscous contributions [Mathai, *et al.* 2018]. As for the domain in which the body's motion occurs and the boundary conditions, let us consider only the Cauchy problem in the whole space (besides, fluid is considered to be at rest at infinity).

Let us note that system of equations (1)-(2) could be presented in the most general form [Novikov & Shmel'tser, 1981], which allows obtaining three integrals of motion for this problem (including integral of energy). Nevertheless, we will consider here only the reduced version of Kelvin–Kirchhoff equations (1)-(2); the aforesaid simplification is applied with respect to the equation of angular momentum conservation (2), assuming 2 principal moments of inertia are equal to each other (in case of *rigid* torus with uniform distribution of density inside its boundaries).

The last but not least, we should especially note that system of equation (1)-(2) is not complete. Indeed, co-rotating frame of a Cartesian coordinate system (fixed in the rotating body) should be transformed to the absolute Cartesian coordinate system via Euler angles or (preferably) Wisdom angles [Ershkov & Shamin, 2018*a*].

## 2. **General presentation of the solution for velocity field.**

The momentum Eqn. (1) of Kelvin–Kirchhoff equations is known to be the system of 3 linear ordinary differential equations (in regard to the time *t*) for 3 unknown functions: $v_1$, $v_2$, and $v_3$ (indeed, Eqns. of the aforementioned type have been considered in [Ershkov, 2017*a*], [Ershkov, 2017*b*], [Ershkov & Shamin, 2018*b*], [Ershkov & Leshchenko, 2019*a*], and [Ershkov & Leshchenko, 2020*c*] accordingly). General presentation of non-homogeneous solution for velocity field has been demonstrated earlier in work [Ershkov & Shamin, 2018*b*].

According to [Kamke, 1971], the system of Eqns. (1) could be considered as having been solved if we obtain a general solution of *the corresponding homogeneous* system (1), $\vec{v}|_0 = \{U, V, W\}$:

$$\frac{d\vec{v}|_0}{dt} + (A\vec{\Omega}) \times \vec{v}|_0 = \vec{0}, \qquad (3)$$

$$\Rightarrow \begin{cases} \dfrac{dU}{dt} = V \cdot (A\Omega_3) - W \cdot (A\Omega_2), \\[1em] \dfrac{dV}{dt} = W \cdot (A\Omega_1) - U \cdot (A\Omega_3), \\[1em] \dfrac{dW}{dt} = U \cdot (A\Omega_2) - V \cdot (A\Omega_1). \end{cases} \qquad (4)$$

The system of Eqns. (3) (or (4)) has *the analytical* way to present the general solution [Ershkov & Shamin, 2018b] in regard to the time *t*:

$$U = -\sigma \cdot \left(\frac{2a}{1+(a^2+b^2)}\right), \quad V = -\sigma \cdot \left(\frac{2b}{1+(a^2+b^2)}\right),$$

$$W = \sigma \cdot \left(\frac{1-(a^2+b^2)}{1+(a^2+b^2)}\right), \qquad (5)$$

here $\sigma = \sigma(x, y, z)$ is some arbitrary (real) function, given by the initial conditions; the real-valued coefficients $a(x,y,z,t)$, $b(x,y,z,t)$ in (5) are solutions of the mutual system of two *Riccati* ordinary differential equations in regard to the time *t*:

$$\begin{cases} a' = \left(\dfrac{A \cdot \Omega_2}{2}\right) \cdot a^2 - (A \cdot \Omega_1 \cdot b) \cdot a - \dfrac{A \cdot \Omega_2}{2}(b^2 - 1) + (A \cdot \Omega_3) \cdot b, \\[1em] b' = -\left(\dfrac{A \cdot \Omega_1}{2}\right) \cdot b^2 + (A \cdot \Omega_2 \cdot a) \cdot b + \dfrac{A \cdot \Omega_1}{2} \cdot (a^2 - 1) - (A \cdot \Omega_3) \cdot a. \end{cases} \qquad (6)$$

Equations (6) above are forming the *Riccati*-type system of ordinary differential equations [Ershkov & Shamin, 2018*b*], [Ershkov & Leshchenko, 2019*a*]. Each of them describes the evolution of function *a* in dependence on the function *b* in regard to the time *t* (and *vice versa*); such a *Riccati* ordinary differential equation has no analytical solution in general case [Kamke, 1971].

Mathematical procedure for checking of the solution (5)-(6) (which is to be valid for the *homogeneous* momentum equations (1) in a form of equations (4)) has been demonstrated in **Appendix A** of [Ershkov & Shamin, 2018*b*], with only the resulting formulae left in the main text.

### 3. <u>Expressions for components of net torque on torus during the motion.</u>

First of all, we should mention the fundamental work [Johnson & Wu, 1979], where the moments of forces acting on a slender torus immersed in creeping (*stationary*, viscous) shear flow were calculated. Furthermore, basing on the aforementioned analytical result, a lot of interesting results in theoretical and numerical findings have been obtained (e.g., see [Thaokar, *et al.* 2007], [Moshkin & Suwannasri, 2012]). These results could be applied to the solving procedure in our article, devoted to the investigating the essentially *non-stationary* case of flow field around the falling torus in *ideal* incompressible fluid, but we should take into account in our analysis that, despite of the assumption of ideal fluid was used at formulation of (1)-(2), the Stokes viscous boundary layer phenomenon on the boundaries of the body has also been assumed. This assumption allows us to use the results of *Johnson & Wu* fundamental work [Johnson & Wu, 1979], in which torus was assumed to be a slim enough to neglect the inertial effects. Meanwhile, in [Johnson & Wu, 1979] motions of a slender torus have been classified into to 5 types (2 types of translational motion, 2 types of rotations as well as the case of expanding torus). We consider here the case of rigid torus only, so the case of expanding torus should be excluded.

As for translational motion of a torus perpendicular to its longitudinal axis, e.g. according to the results of [Thaokar, *et al.* 2007], torques are felt only *locally* by the torus's surface, which means that net torque about its center-line is zero for this type of translational motion (in case of a rotating torus) due to the symmetry of the flow.

If we consider the case of broadwise translation of a rotating torus in the outer flow (i.e., translation along the longitudinal axis of a torus), the total dimensional net torque on the torus about the center-line [Thaokar, *et al.* 2007] should be given by

$$T_z(v_3) = 4\pi^2 \rho_f \cdot \nu \cdot d^2 \cdot v_3 \cdot \left( \frac{2\ln(8/\varepsilon) - 1}{2\ln(8/\varepsilon) + 1} \right) \qquad (7)$$

here, expression (7) is the central result of work [Johnson & Wu, 1979] ($\varepsilon = d/D$); but this type of torques (due to translational motion) is felt also only *locally* by the torus's surface, that's why the net torque about its centerline is zero as well (for the rotating torus). Meanwhile, we should note that formulae (67)-(68) in [Thaokar, *et al.* 2007] differ from appropriate *Johnson & Wu* formulae, see expression (28) in [Johnson & Wu, 1979]: indeed, while expression for torque had been presented per unit of line about the torus center-line in [Johnson & Wu, 1979], the corresponding expression in [Thaokar, *et al.* 2007] was used as the dimensional net torque on the torus about the center-line (but in [Thaokar, *et al.* 2007] was used function $\log(8/\varepsilon)$ instead of $\ln(8/\varepsilon)$ in [Thaokar, *et al.* 2007]; as we know, $\log(8/\varepsilon) \cong 0{,}434 \cdot (8/\varepsilon)$, so, these coefficients essentially differ from each other).

In case of motion of the spinning torus (or the torus rotating about its axis of symmetry), the total 2-side dimensional net torque on the torus about the center-line [Thaokar, *et al.* 2007] is given by ($\varepsilon = d/D$):

$$T_3 = -4\pi^2 \rho_f \cdot \nu\, d^2 \cdot D \cdot \Omega_3 \cdot \frac{(6\ln(8/\varepsilon) - 13)}{4(\ln(8/\varepsilon) - 2)} \tag{8}$$

The last but not least, we should mention the case of edge rotation of a torus; namely, rotation takes place in a plane which perpendicular to the rotation about its axis of symmetry, whereas the net torque (which should be integrated out over the surface of torus about its center-line) is given according to the results of work [Goren & O'Neill, 1980]

$$T_{1,2} = -g_{x,y} \cdot 8\pi \rho_f \cdot \nu \cdot D^3 \cdot \Omega_{1,2} \tag{9}$$

here, we should use the sign "-" before the expression the net torque (because authors of fundamental work [Goren & O'Neill, 1980] had been concentrated on the obtaining the absolute magnitude of the appropriate components of net torque on the surface of torus); obviously, only negative components of net torque should decrease the appropriate components of angular velocity of rotation during the motion of a torus in the flows of the viscous fluid about its surface.

As for the data for the absolute magnitudes of the constants $g_{x,y}$ (dimensionless torque coefficients), we should refer to the Table 1 in [Goren & O'Neill, 1980]:

Table 1. Values of dimensionless torque coefficients for torus of various geometry ($d \neq 0$, $(D/d) \neq 1$)

| (D/d) - 1 | 1.01 | 1.1 | 1.2 | 1.4 | 1.6 | 1.8 | 2 | 3 | 4 |
|---|---|---|---|---|---|---|---|---|---|
| $g_x = g_y$ | 0.6485 | 0.6362 | 0.6238 | 0.6027 | 0.5855 | 0.5711 | 0.5590 | 0.5184 | 0.4943 |

| (D/d) - 1 | 5 | 6 | 8 | 10 | 20 | 40 | 60 | 80 | 100 |
|---|---|---|---|---|---|---|---|---|---|
| $g_x = g_y$ | 0.4776 | 0.4646 | 0.4451 | 0.4304 | 0.3857 | 0.3435 | 0.3206 | 0.3055 | 0.2943 |

## 4. Solutions for regime of a spinning torus in case of laminar flow (at slow regime of torus rotation).

First, we can write out the solving procedure for angular momentum equations (2) in a simple case of spinning torus (including rotation of a torus about its axis of symmetry). It means that we consider the 3D rotations of a torus (as a rigid body), which is also spinning about its axis of symmetry. First of all, we should note that (2) is the system of 3 nonlinear differential equations (10) with respect to $\vec{K} = \{I_i \cdot \Omega_i\}$ (with all coefficients depending on time $t$):

$$\frac{d\vec{K}}{dt} + [\vec{\Omega} \times \vec{K}] = \vec{T}, \quad \Rightarrow \quad \begin{cases} \dfrac{dK_1}{dt} = K_2 \cdot (\Omega_3) - K_3 \cdot (\Omega_2) + T_1, \\[2mm] \dfrac{dK_2}{dt} = K_3 \cdot (\Omega_1) - K_1 \cdot (\Omega_3) + T_2, \\[2mm] \dfrac{dK_3}{dt} = K_1 \cdot (\Omega_2) - K_2 \cdot (\Omega_1) + T_3. \end{cases} \qquad (10)$$

$$\Rightarrow \begin{cases} \dfrac{dK_1}{dt} = K_2 \cdot \left(\dfrac{K_3}{I_3}\right) - K_3 \cdot \left(\dfrac{K_2}{I_2}\right) + T_1, \\ \dfrac{dK_2}{dt} = K_3 \cdot \left(\dfrac{K_1}{I_1}\right) - K_1 \cdot \left(\dfrac{K_3}{I_3}\right) + T_2, \\ \dfrac{dK_3}{dt} = K_1 \cdot \left(\dfrac{K_2}{I_2}\right) - K_2 \cdot \left(\dfrac{K_1}{I_1}\right) + T_3. \end{cases} \Rightarrow \begin{cases} \dfrac{dK_1}{dt} = K_2 \cdot K_3 \cdot \left(\dfrac{I_2 - I_3}{I_2 \cdot I_3}\right) + T_1, \\ \dfrac{dK_2}{dt} = K_1 \cdot K_3 \cdot \left(\dfrac{I_3 - I_1}{I_1 \cdot I_3}\right) + T_2, \quad (11) \\ \dfrac{dK_3}{dt} = K_1 \cdot K_2 \cdot \left(\dfrac{I_1 - I_2}{I_1 \cdot I_2}\right) + T_3. \end{cases}$$

Assuming 2 principal moments of inertia are equal to each other ($I_1 = I_2$), we obtain from 3-rd equation of system (11)

$$\frac{dK_3}{dt} = T_3 \Rightarrow K_3 \equiv I_3 \cdot \Omega_3 = \int (T_3) dt \qquad (12)$$

Thus, the key governing factors (which should determine the dynamics of a torus rotation in a fluid) are the 3 components of torque $\vec{T}$, besides, two of them should be equal to zero for the symmetric torus $\{T_1, T_2\} = 0$.

That's why regimes of torus rotation in a fluid obviously depends on the expressions for the component of torque $T_3$. Moreover, the variability observed in the regimes of rotation during falling of body in a form of slim torus can be associated with two basic regimes of flows, laminar and turbulent (along with switching between them).

Let us discuss the case of laminar flow at slow regime of torus rotation, $\Omega_3 \neq const$. Taking into account that

$$I_1 = (m/4) \cdot ((D^2/2) + (5d^2/8)), \quad I_3 = (m/4) \cdot ((D^2/2) + (3d^2/4)),$$

we obtain from (11)-(12), $\Omega_3(0) = const$:

$$I_3 \cdot \frac{d\Omega_3}{dt} = -4\pi^2 \rho_f \cdot v \cdot d^2 \cdot D \cdot \Omega_3 \cdot \frac{(6\ln(8/\varepsilon) - 13)}{4(\ln(8/\varepsilon) - 2)} \Rightarrow$$

$$\Omega_3(t) = \Omega_3(0) \cdot \exp\left(-\omega_3 \cdot t\right) \tag{13}$$

$$\left\{ \omega_3 = \frac{16\pi^2 \rho_f \cdot v \cdot d^2 \cdot D \cdot \frac{(6\ln(8/\varepsilon) - 13)}{(\ln(8/\varepsilon) - 2)}}{m \cdot (2D^2 + 3d^2)} \right\}$$

Let us multiply the 1-st equation of system (11) on $K_1$, the 2-nd equation on $K_2$, then sum these transformed equations one to another (note also that $I_1 = I_2$, whereas $T_1 = T_2 = 0$); we obtain:

$$K_1 \cdot \frac{dK_1}{dt} + K_2 \cdot \frac{dK_2}{dt} = 0 \Rightarrow,$$

$$K_1^2 + K_2^2 = K_0^2 = const \Rightarrow K_2 = \pm\sqrt{K_0^2 - K_1^2} \tag{14}$$

Now, as for the 1-st equation of system (11), let us substitute expression for $K_2$ from Eqn. (14); it yields as below

$$\frac{dK_1}{dt} = \pm\sqrt{K_0^2 - K_1^2} \cdot K_3 \cdot \left(\frac{I_1 - I_3}{I_1 \cdot I_3}\right), \Rightarrow \left\{ \int \frac{dK_1}{\sqrt{K_0^2 - K_1^2}} = \arcsin\left(\frac{K_1}{|K_0|}\right) \right\}$$

$$\arcsin\left(\frac{K_1}{|K_0|}\right) = \mp\left(\frac{(m/4) \cdot (d^2/8)}{(m/4) \cdot ((D^2/2) + (5d^2/8))}\right) \cdot \Omega_3(0) \cdot \int \exp\left(-\omega_3 \cdot t\right) dt, \Rightarrow$$

$$\Rightarrow \Omega_1 = \pm\omega_1 \cdot \sin\left(\frac{d^2}{(4D^2 + 5d^2)} \frac{\Omega_3(0)}{\omega_3} \cdot \exp\left(-\omega_3 \cdot t\right)\right) \tag{15}$$

$$\left\{ \omega_1 = \frac{32|K_0|}{m \cdot (4D^2 + 5d^2)}, \quad \omega_3 = \frac{16\pi^2 \rho_f \cdot v \cdot d^2 \cdot D \cdot \frac{(6\ln(8/\varepsilon) - 13)}{(\ln(8/\varepsilon) - 2)}}{m \cdot (2D^2 + 3d^2)} \right\}$$

Let us schematically imagine at Figs.1-2 the component $\Omega_1$ of angular velocity (15) in regard to the time $t$ (where we designate $x = t$ just for the aim of presenting the plot of solution):

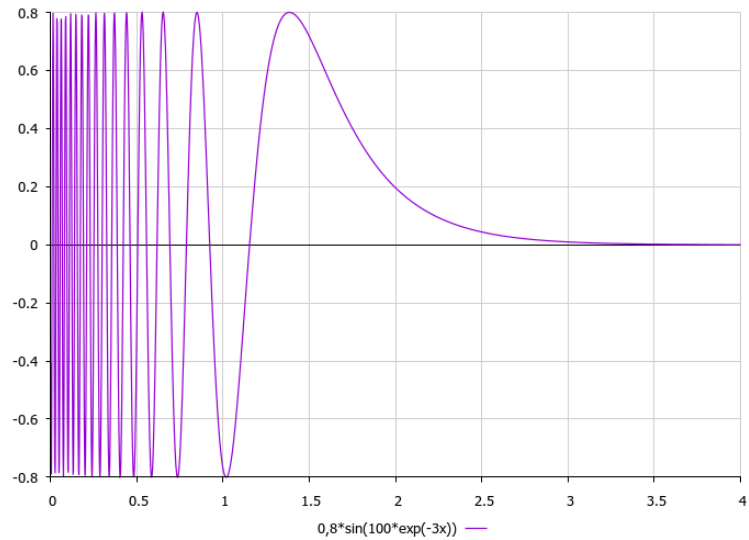

Fig.1. A *schematic* plot of the component $\Omega_1(t)$ of angular velocity (15).

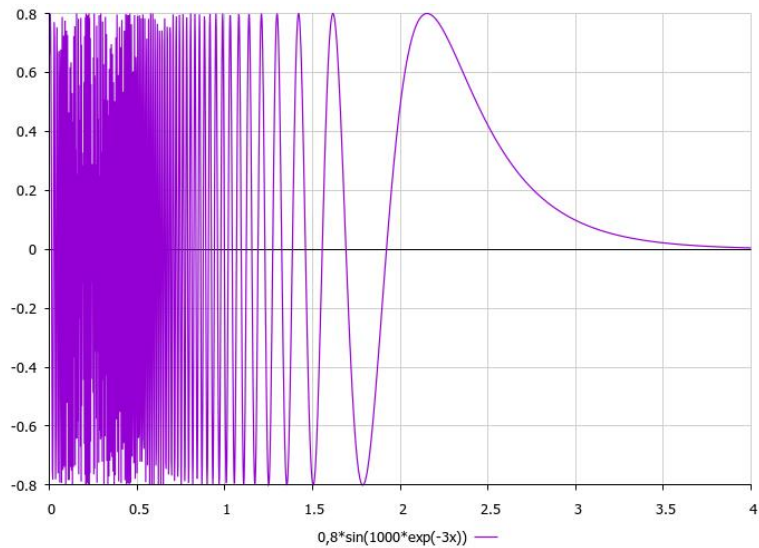

Fig.2. A *schematic* plot of the component $\Omega_1(t)$ of angular velocity (15).

Thus, we have fully solved the system of equations (2) in case (8) of laminar flow around the spinning torus at slow regime of torus rotation when 2 principal moments of inertia are equal to each other ($I_1 = I_2$).

## 5. Solutions for regime of arbitrary rotating torus in case of laminar flow about the torus during rotation.

Now, we can illuminate or clarify the solving procedure for angular momentum equations (2) in a case of the arbitrary rotating torus (not only a spinning torus about its axis of symmetry) at slow regime of torus rotation. Taking into account expressions (8)-(9) the appropriate components of net torque on the surface of torus, we obtain from (11)-(12):

$$\Rightarrow \begin{cases} I_1 \cdot \dfrac{d\Omega_1}{dt} = \Omega_2 \cdot \Omega_3 \cdot (I_1 - I_3) - g_y \cdot 8\pi \rho_f \cdot v \cdot D^3 \cdot \Omega_1, \\ \\ I_1 \cdot \dfrac{d\Omega_2}{dt} = \Omega_1 \cdot \Omega_3 \cdot (I_3 - I_1) - g_y \cdot 8\pi \rho_f \cdot v \cdot D^3 \cdot \Omega_2. \end{cases} \quad (16)$$

Let us multiply the 1-st equation of system (16) on $\Omega_1$, the 2-nd equation on $\Omega_2$, then sum these transformed equations one to another; we obtain:

$$I_1 \cdot \Omega_1 \cdot \frac{d\Omega_1}{dt} + I_1 \cdot \Omega_2 \cdot \frac{d\Omega_2}{dt} = - g_y \cdot 8\pi \rho_f \cdot v \cdot D^3 \cdot (\Omega_1^2 + \Omega_2^2) \quad \Rightarrow \quad ,$$

$$\Omega_1^2 + \Omega_2^2 = \Omega_0^2 \cdot \exp\left(- \frac{g_y \cdot 16\pi \rho_f \cdot v \cdot D^3}{I_1} \cdot t \right), \quad \Omega_0^2 = const \quad \Rightarrow$$

$$\Omega_2 = \pm \sqrt{\Omega_0^2 \cdot \exp\left(- \frac{g_y \cdot 16\pi \rho_f \cdot v \cdot D^3}{I_1} \cdot t \right) - \Omega_1^2} \quad (17)$$

whereby we conclude that the definite restriction is valid for the component of angular velocity $\Omega_1$

$$|\Omega_1| \leq |\Omega_0| \cdot \exp\left(-\frac{g_y \cdot 8\pi \rho_f \cdot \nu \cdot D^3}{I_1} \cdot t\right) \qquad (18)$$

Now, as for the 1-st equation of system (16), let us substitute expression for $\Omega_2$ from Eqn. (17); it yields as below

$$I_1 \cdot \frac{d\Omega_1}{dt} = \pm\left(\sqrt{\Omega_0^2 \cdot \exp\left(-\frac{g_y \cdot 16\pi \rho_f \cdot \nu \cdot D^3}{I_1} \cdot t\right) - \Omega_1^2}\right) \cdot \Omega_3 \cdot (I_1 - I_3) - g_y \cdot 8\pi \rho_f \cdot \nu \cdot D^3 \cdot \Omega_1 \quad (19)$$

where expression for $\Omega_3$ is given by (13); $I_1 = (m/4) \cdot ((D^2/2) + (5d^2/8))$, $I_3 = (m/4) \cdot ((D^2/2) + (3d^2/4))$. Obviously, ordinary differential equation (19) of the 1-st order could be solved only by numerical methods only. We should also note that in view of a priori estimate (18), the qualitative properties of this equation's solution are close to the properties of a linear one (but, nevertheless, we consider the mathematical procedure of presenting the approximate solution in **Appendix**).

Thus, we have fully solved the system of equations (2) in case (8)-(9) of laminar flow around the arbitrary rotating torus at slow regime of rotation of the symmetrical torus ($I_1 = I_2$).

## 6. <u>Discussion.</u>

As we can see from derivation above, system of Kelvin-Kirchhoff equations (which governs by the dynamics of body's motion in an ideal fluid) is proved to be very hard to solve analytically even in simple case of *falling* the rotating torus.

Indeed, at first step we should solve the angular momentum equation (2) (with the main aim to obtain the components of angular velocity $\{\Omega_i\}$ of rotation for the falling body). In the aforementioned solving procedure, we take into consideration that the Stokes viscous boundary layer phenomenon on the boundaries of the body has also been assumed at formulation of basic Kelvin-Kirchhoff equations (1)-(2). It allows us to use the results of *Johnson & Wu* work [Johnson & Wu, 1979] in formulae (10)-(15), in which analytical solution for components of angular velocity of rotations for the spinning torus was obtained (as well as we can use results of *Goren & O'Neill* work [Goren & O'Neill, 1980] in formulae for the components of a net torque on the surface of torus, in case of edge rotation of a torus).

Just for example, as additional assumption in the aforementioned solving procedure, we assume (8) for the component of fluid's torque vector $T_3$ (in case of laminar flow at slow regime of the spinning torus rotation).

Then at 2-nd step we should solve *homogeneous* momentum equation (1) in a form (4) in regard to the time *t* (depending on 3 components of angular velocity $\vec{\Omega}$ of torus rotation (13)-(15)). In general case, it is almost impossible for the reason that each component of $\vec{\Omega}$ depends quasi-periodically or even chaotically on time *t* (for example, in case of turbulent flow at torus rotation).

At the next 3-rd step we should solve the *non-homogeneous* momentum equation (1) with the *non-homogeneous* part of solution for velocity field $\vec{v}$ (depending on the *homogeneous* part of solution for velocity field $\vec{v}|_0 = \{U, V, W\}$, as well as depending on the components of fluid force in expression for $\vec{f}$, Eqns. (1)). Meanwhile, we should take also into account in expression for the external fluid forces $\vec{f}$ that the torus is supposed to be under the action of additional force which is associated (and components of which are directly proportional) with to the net inward-outward circulation of the fluid's flow via the central perforation in a torus during its motion in a fluid [Bryan, 1893] (the aforementioned total circulation is proportional to the volume of liquid per unit time flowing through aperture or across the central perforation relatively to the torus, with inverse coefficient of proportionality equals to the density of fluid). The direction of such the additional force is fixed in the appropriate position relative to the solid body during its motion in a fluid [Bryan, 1893], preferably along the axis of symmetry of torus rotation. Overall, this phenomenon could be considered as stabilizing factor acting along all the trajectory of falling of the rotating torus deep in the incompressible fluid.

Thus, we have fully solved the system of equations in case of laminar flow around the arbitrary rotating torus at slow regime of torus rotation.

Let us list below the obvious assumptions (or simplifications) which have been used at the formulation or the solution of the problem:

- the assumption of ideal fluid was used (it means that the effects of viscous fluid are preferably ignored, excepting Stokes boundary layer phenomenon on the boundaries of the body with Reynolds number much less than << 1 - the possibility of this approach application to the motion regime, which is far from periodic one, is a debatable question depending on the range of Reynolds numbers for a possible physical interpretation of the obtained results);

- fluid is assumed to be incompressible (density is supposed to be constant);
- rotation of torus is assumed to be *rigid* which means that distances between various points inside the body should not be elongated significantly;
- the shape of torus is assumed to be very close to symmetric form, $I_1 = I_2$;
- fluid flow enveloping the torus is assumed to be potential (as it has been formulated initially in [Kirchhoff, 1877] for the motions of rigid bodies of various forms [Ershkov, *et al.* 2020*a*], [Ershkov, *et al.* 2020*b*], which have been described by Kelvin–Kirchhoff equations);
- we consider only the Cauchy problem in the whole space (besides, fluid is considered to be at rest at infinity).

Ending discussion, we should note that we can see from the aforementioned list of simplifications that the real conditions for falling or ascending bodies in fluids (e.g., falling the coin in the water) are definitely very far from the ideal formulation of the Kelvin–Kirchhoff problem. As for developing the main idea, the suggested approach can be used in the future researches for analysis of the falling or ascending motions in fluids of other rigid or *quasi-rigid* bodies with symmetry of rotation (e.g., in a form of slim disc), not only in simple case of rigid sphere or in case of *quasi-rigid* spheroid. Predicting of regimes of forced oscillations is obviously useful in technical applications.

## 7. Conclusion.

We have presented in this communication a new solving procedure for Kelvin–Kirchhoff equations, considering the dynamics of falling the *rigid* rotating torus in an ideal incompressible fluid, assuming additionally the dynamical symmetry of rotation for the rotating body, $I_1 = I_2$.

Fundamental law of angular momentum conservation has been used for the aforementioned solving procedure. The system of *Euler* equations for dynamics of torus rotation has been explored in regard to the existence of an analytic way of presentation for the approximated solution (where we consider the case of laminar flow at slow regime of torus rotation). The second finding is associated with consideration of that the Stokes viscous boundary layer phenomenon on the boundaries of the body has also been assumed at formulation of basic Kelvin-Kirchhoff equations. E.g., it allows us to use e.g. the results of *Johnson & Wu* fundamental work [Johnson & Wu, 1979], in which analytical expression for the component of fluid's torque vector $T_3$ was obtained (as well as we can use results of *Goren & O'Neill* work [Goren & O'Neill, 1980]). The results of calculations for the components of angular velocity $\{\Omega_i\}$ should then be used for solving momentum equation of Kelvin–Kirchhoff system. Thus, the full system of equations of Kelvin–Kirchhoff problem has been explored with respect to the existence of an analytic way of presentation of the exact solution.

Also, remarkable articles [Miloh & Landweber, 1981], [Galper & Miloh, 1994], [Chernousko, *et al.* 2017], [Routh, 1905], [Yehia, 1986] and [Lamb, 1895] should be cited, which concern the generalization of the problem under consideration. In [Chernousko, *et al.* 2017], see paragraph 5.2 and formulae (5.20)-(5.21) with respect to the components of the resistance torque vector $\{T_i\}$ in a fluid (which are often considered to be proportional to the appropriate components of angular velocity $\{\Omega_i\}$, multiplied by the appropriate the principal central moments of inertia of the rigid body). Retrospectively, as far as we know, the cases of components of air's torque vector $\{T_i\}$ being approximately proportional to the appropriate components of angular velocity $\{\Omega_i\}$ have been first considered in paragraphs 147 and 202d of work [Routh, 1905]. In [Yehia, 1986], the full similarity between equations of motion of a rigid body in an ideal incompressible fluid and, from other hand, the equations of motion of a gyrostat about fixed point (under the influence of both the potential and gyroscopic forces) was established.

The last but not least, let us also remind that cases of motion of the perforated body have been considered in [Yehia, 1986], [Lamb, 1895] from theoretical point of view (i.e., motion of rigid body bounded by a multi-connected surface in an infinite ideal incompressible fluid). Obviously, the total net torque caused by the multi-connectedness in case of the torus surface (such the net torque should be integrated out over the surface of torus about its center-line) vanishes to zero due to the symmetry of rotation during the motion in a viscous incompressible fluid (the analogue of the appropriate torus dynamics was considered in [Moshkin & Suwannasri, 2012]).

## 8. <u>Acknowledgements.</u>

classes - zigzagging, helical spiral motion, and the chaotic regime of oscillations.

Based on these experiments, we can confirm the conclusion made by authors in Section 3.3 of work [Vetchanin & Gladkov, 2012] that in most cases the torus appears to be reach the bottom in a plane position indeed (with respect to the bottom).

## Conflict of interest

Authors declare that there is no conflict of interests regarding publication of article.

Remark regarding contributions of authors as below:

In this research, Dr. Sergey Ershkov is responsible for the general ansatz and the solving procedure, simple algebra manipulations, calculations, results of the article in Sections 1-6 and also is responsible for the search of exact solutions.

Dr. Dmytro Leshchenko is responsible for theoretical investigations as well as for the deep survey in literature on the problem under consideration.

Ayrat Giniyatullin is responsible for testing the initial conditions for the approximated solutions (as well as is responsible for the plots and graphical solutions).

All authors agreed with the results and conclusions of each other in Sections 1-7.

## Appendix (approximate solving of Eqn. (19)).

Taking into account that we consider the case of *slow* regime of torus rotation (which stems from (18)), we obtain from (19) by series of Taylor expansions as 1-st non-linear approximation:

$$I_1 \cdot \frac{d\Omega_1}{dt} = \pm |\Omega_0| \cdot \exp\left(-\frac{g_y \cdot 8\pi \rho_f \cdot \nu \cdot D^3}{I_1} \cdot t\right) \cdot \left(1 - \frac{\Omega_1^2}{2\Omega_0^2 \cdot \exp\left(-\frac{g_y \cdot 16\pi \rho_f \cdot \nu \cdot D^3}{I_1} \cdot t\right)}\right) \cdot \Omega_3 \cdot (I_1 - I_3) -$$

$$- g_y \cdot 8\pi \rho_f \cdot \nu \cdot D^3 \cdot \Omega_1, \quad \Rightarrow$$

$$\frac{d\Omega_1}{dt} = \mp \left\{\exp\left(\left(\frac{g_y \cdot 8\pi \rho_f \cdot \nu \cdot D^3}{I_1} - \omega_3\right) \cdot t\right) \cdot \left(\frac{\Omega_3(0)}{2|\Omega_0|}\right) \cdot \frac{(I_1 - I_3)}{I_1}\right\} \cdot \Omega_1^2 - \left(\frac{g_y \cdot 8\pi \rho_f \cdot \nu \cdot D^3}{I_1}\right) \cdot \Omega_1 \quad (20)$$

$$\pm |\Omega_0| \cdot \exp\left(-\left(\frac{g_y \cdot 8\pi \rho_f \cdot \nu \cdot D^3}{I_1} + \omega_3\right) \cdot t\right) \cdot \Omega_3(0) \cdot \frac{(I_1 - I_3)}{I_1},$$

where equation (20) above is the *Riccati*-type ODE of the 1-st order [Kamke, 1971], which has no analytical solution in general case; it is worth also to note that due to special character of the solutions of *Riccati*-type ODEs [Ershkov, 2017*a*], [Ershkov, 2017*b*], [Ershkov & Shamin, 2018*b*], [Ershkov & Leshchenko, 2019*b*], and [Ershkov & Shamin, 2020d] there is the possibility for sudden *jumping* in the magnitude of the solution at some time $t_0$. In the physical sense, such the aforementioned jumping of *Riccati*-type solutions could be associated with the effect of a sudden acceleration/deceleration of the angular velocity (e.g., the component $\Omega_1(t)$) at a definite moment of parametric time $t_0$.